\documentclass[11pt]{article}
\usepackage{fleqn,cospar}

\usepackage[dvips]{color}

\usepackage{url}

\usepackage{graphicx}
\usepackage[figuresright]{rotating}

\newcommand\mdot   {\hbox {${\dot M}$}}

\newcommand\nubreak {$\nu_{\rm Break}$}
\newcommand\nuqpo   {$\nu_{\rm QPO}$}

\newcommand\medd    {${\dot M}_{\rm Edd}$}

\newcommand\nubreakbold {\boldmath{$\nu_{\rm BREAK}$}}
\newcommand\nuqpobold   {\boldmath{$\nu_{\rm QPO}$}}

\title{Quasi-periodic oscillations and noise in neutron star and
black-hole X-ray binaries}

\author{Rudy Wijnands\footnote{Chandra Fellow}\address{Center for
Space Research, Massachusetts Institute of Technology, 77
Massachusetts Avenue, Cambridge, MA 02139, USA}}

\begin{document}

\author{Rudy Wijnands\footnote{Chandra Fellow}\address{Center for
Space Research, Massachusetts Institute of Technology, 77
Massachusetts Avenue, Cambridge, MA 02139, USA}}

\maketitle

\begin{abstract}
Before the launch of the Rossi X-ray Timing Explorer (RXTE) satellite,
the differences in the rapid X-ray variability between the two main
types of neutron star binaries (i.e., the Z and atoll sources) could
be explained by invoking different mass accretion rates and magnetic
field strengths.  However, the results obtained with RXTE now show
that these systems are more similar than previously thought and
although differences in mass accretion rate are still likely, the
differences in the magnetic field strength have become questionable.
The great similarities between the neutron star systems and the
black-hole candidates at low mass accretion rates also point towards a
similar origin of their timing phenomena indicating that the presence
or absence of a solid surface, a magnetic field, or an event horizon
do not play a significant role in the production mechanisms for the
rapid X-ray variability.

\end{abstract}

\section*{NEUTRON STAR SYSTEMS: THE PRE-RXTE VIEW}

\begin{figure}[t]
\begin{center}
\includegraphics[width=110mm]{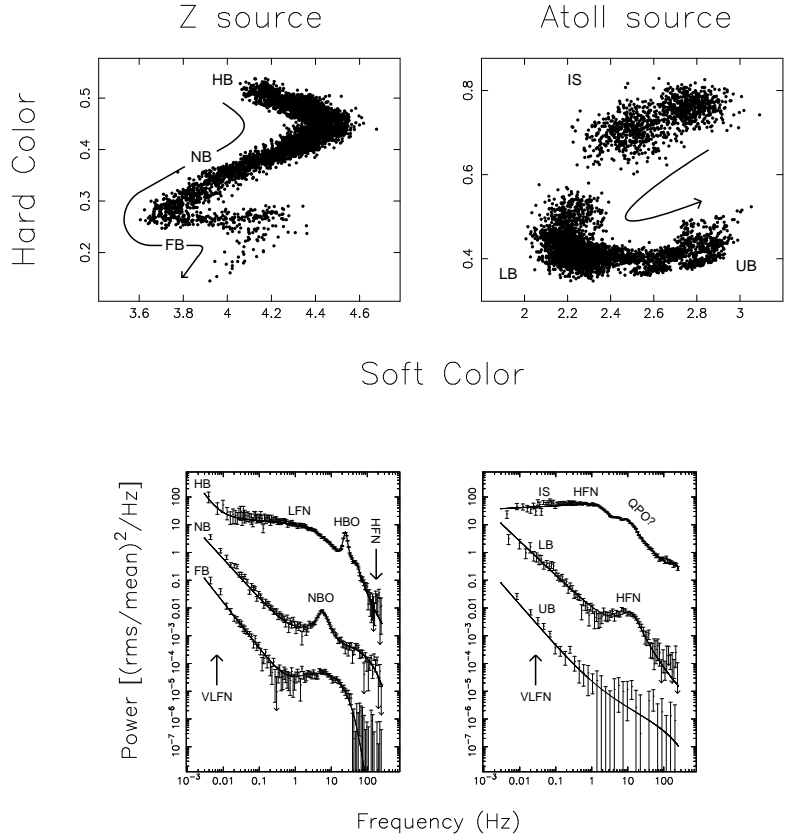}
\end{center}
\caption{X-ray color-color diagram ({\it top}) and power spectrum
({\it bottom}) typical of Z ({\it left}) and atoll sources ({\it
right}). The direction of \mdot~is indicated by the arrows in the CDs.
The power spectra of the HB and the IS are shifted upwards by a factor
of $10^4$; the power spectra of the NB and the LB by a factor of
$10^2$ (Wijnands 1999). \label{fig:z_atoll} }
\end{figure}

Before the start of the RXTE mission, the neutron star (NS) low-mass
X-ray binaries (LMXBs) had been intensively studied with previous
X-ray instruments (i.e., EXOSAT and Ginga). The introduction of X-ray
color-color diagrams (CDs; e.g., Hasinger \& van der Klis 1989) proved
to be extremely useful for the study of the correlations between the
changes in X-ray spectrum and X-ray timing behavior of these systems.
On the basis of this correlated behavior, they were classified into
the Z sources and the atoll sources (Hasinger \& van der Klis
1989). The Z sources trace out a Z shaped track in the CD (see
Fig.~\ref{fig:z_atoll}) with the branches labeled, from top to bottom,
the horizontal branch (HB), the normal branch (NB), and the flaring
branch (FB). The power spectra show (Fig.~\ref{fig:z_atoll}) on the
horizontal branch, strong band-limited noise (called low frequency
noise or LFN) which cuts off below several Hertz, simultaneous with
15--60 Hz quasi-periodic oscillations (QPOs), which are called
horizontal branch oscillations (HBOs). On the normal branch these QPOs
can still be seen, often simultaneous with other 5--7 Hz QPOs, which
are called normal branch oscillations (NBOs).  The NBOs smoothly merge
with the 7--20~Hz QPOs seen on the flaring branch, the flaring branch
oscillations (FBOs). On all branches also two other noise components
are found, one at very low frequencies (the very low frequency noise
or VLFN), following a power law, and one at frequencies above 10 Hz
(the high frequency noise or HFN), which cuts off between 50 and 100
Hz. Motion of the source along the Z track is thought (e.g., Hasinger
\& van der Klis 1989) to be due to variations in the mass accretion
rate (\mdot), which is lowest on the horizontal branch, increasing on
the normal branch onto the flaring branch, where it reaches the
Eddington limit \mdot~(\medd).  The atoll sources trace out a curved
branch in the CD (Fig.~\ref{fig:z_atoll}), which can be divided in the
island state (IS) and the banana branch (sub-divided into the lower
banana [LB] and upper banana [UB] branch). The power spectrum in the
island state (Fig.~\ref{fig:z_atoll}) is dominated by very strong
(sometimes more than 20\% rms amplitude) band-limited noise,
superimposed on which are broad bumps (sometimes called QPOs).  The
band-limited noise is also called HFN but it is at much lower
frequencies then the HFN observed in the Z sources and most likely
they are not related. In the power spectrum on the banana branch only
a weak (several percent rms amplitude) power law noise component at
low frequencies is observed (the VLFN), with sometimes another weak
(again several percent rms) noise component at higher frequencies (the
HFN) which cuts off around 10 Hz.  For atoll sources \mdot~is thought
(e.g., Hasinger \& van der Klis 1989) to be lowest in the IS, and
increasing on the banana branch from the lower banana onto the upper
banana branch.

Prompted by the Z-atoll source phenomenology and the QPO models (e.g.,
Alpar \& Shaham 1985; Lamb et al. 1985; Fortner et al. 1989; Alpar et
al. 1992) it was proposed (e.g., Hasinger \& van der Klis 1989) that
the differences between the Z and the atoll sources could be explained
by assuming that the Z sources can reach higher mass accretion rates
than the atoll sources and that the neutron stars in the Z sources
have a higher magnetic field strength $B$ than the neutron stars in
the atoll sources.  The \mdot~difference could explain the luminosity
difference and the presence of the N/FBO in the Z sources; the $B$
difference could explain the presence of the HBO in the Z sources (see
van der Klis 1995 and references therein). This hypothesis predicted
that NBO-like QPOs could be present in the atoll source if they would
reach \medd, but because of their lower $B$, HBO-like QPOs were not
expected in the atoll sources, or at least with a much weaker
amplitude than the HBOs observed in the Z sources (van der Klis 1995;
however, it is possible, although not likely, that a connection exist
between \mdot~and $B$ in such a way that HBO-like QPOs might be
possible in the atoll sources at comparable strength as the HBOs in
the Z sources). Below I will demonstrate that both predictions were
proven to be incorrect.

\section*{NEUTRON STAR SYSTEMS: THE RXTE-ERA}

The RXTE satellite fulfilled its promise and produced (and still is
producing) a wealth of new data on neutron star X-ray
binaries. Important discoveries such as those of the first
accretion-driven millisecond X-ray pulsar SAX J1808.4--3658 (Wijnands
\& van der Klis 1998a), the nearly-coherent oscillations during type I
X-ray bursts (e.g., Strohmayer et al. 1996), and the two simultaneous
QPOs at frequencies between 300 and 1300 Hz (the twin kHz QPOs; e.g.,
Strohmayer et al. 1996; van der Klis et al. 1996) considerably
improved our knowledge of those systems. I will not discuss those
phenomena (see van der Klis 2000 for an excellent review) but instead
I will concentrate on the timing phenomena below a few hundred Hertz.

\subsection*{HBO- and NBO-like QPOs in the atoll sources: Questioning
QPO models}

\begin{figure}[t]
\begin{center}
\includegraphics[width=150mm]{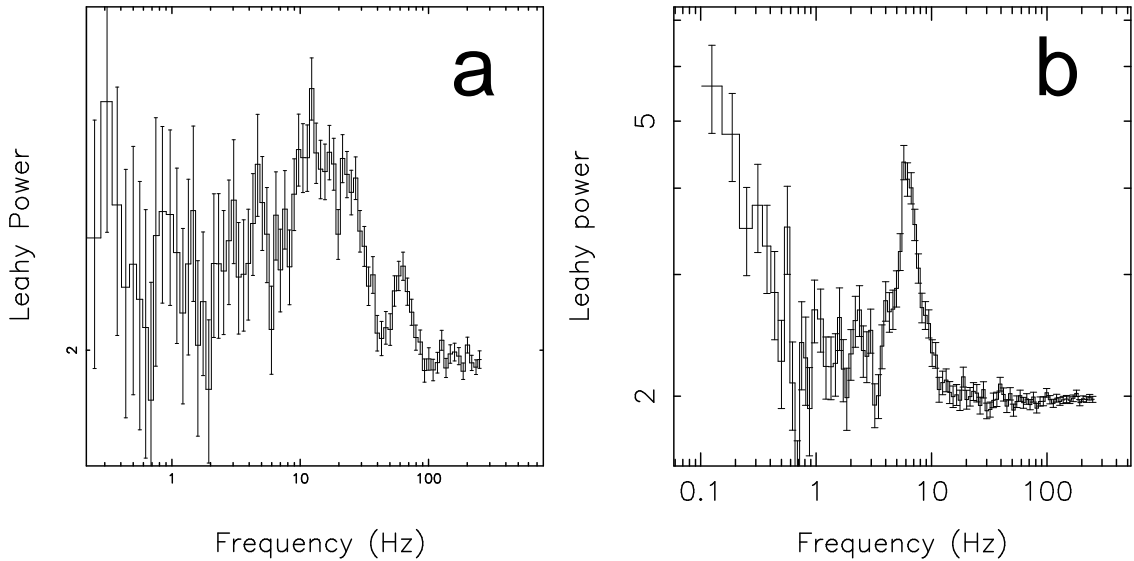}
\end{center}
\caption{The 63 Hz QPO found in the atoll source Serpens X-1 ({\it a})
and the 7 Hz QPO found in the atoll source 4U 1820--30 ({\it b};
Wijnands et al. 1999a).
\label{fig:atoll_qpos} }
\end{figure}

Soon after the launch of {\it RXTE}, it became clear that also the
atoll sources can exhibit 20--70 Hz QPOs (e.g., Strohmayer et
al. 1996; Ford \& van der Klis 1998; Fig.~\ref{fig:atoll_qpos}{\it
a}). The QPOs are usually seen together with the HFN
(Fig.~\ref{fig:atoll_qpos}{\it a}) and are clearly related to the
broad bump seen in the IS (e.g., Ford \& van der Klis 1998). The
frequencies of these low-frequency QPOs are correlated with the kHz
QPOs frequencies, similar to what has been found for the HBOs and the
kHz QPOs in the Z sources (e.g., Ford \& van der Klis 1998; see van
der Klis 2000). The similarities between these types of low-frequency
QPOs indicates that it is likely that they are physically related to
each other (however, the picture is not yet fully understood as
discussed below).  The presence of such QPOs in these sources (and at
similar strength as the HBOs; Ford \& van der Klis 1998) question the
validity of the proposed higher $B$ in the Z sources compared to the
atoll sources.

Besides the 20--70 Hz QPOs, in two atoll sources (4U 1820--30 and the
possible atoll source XTE J1806--246), QPOs were also found with a
frequency of $\sim$7 Hz when those two sources were accreting at their
highest observed \mdot~(i.e., the upper banana branch in 4U 1820--30;
Wijnands \& van der Klis 1999c; Wijnands et al. 1999a). These QPO
properties are similar to the NBOs observed in the Z sources. If these
QPOs are indeed physically related then models explaining the NBOs
(e.g., Fortner et al. 1989; Alpar et al. 1992), which require
near-Eddington mass accretion rates, will not hold. The formation
mechanism behind these QPOs is already activated in 4U 1820--30 well
below \medd~(Wijnands et al. 1999a).

\subsection*{The 1 Hz QPOs in the X-ray dippers and SAX J1808.4--3658}

\begin{figure}[t]
\begin{center}
\includegraphics[width=150mm]{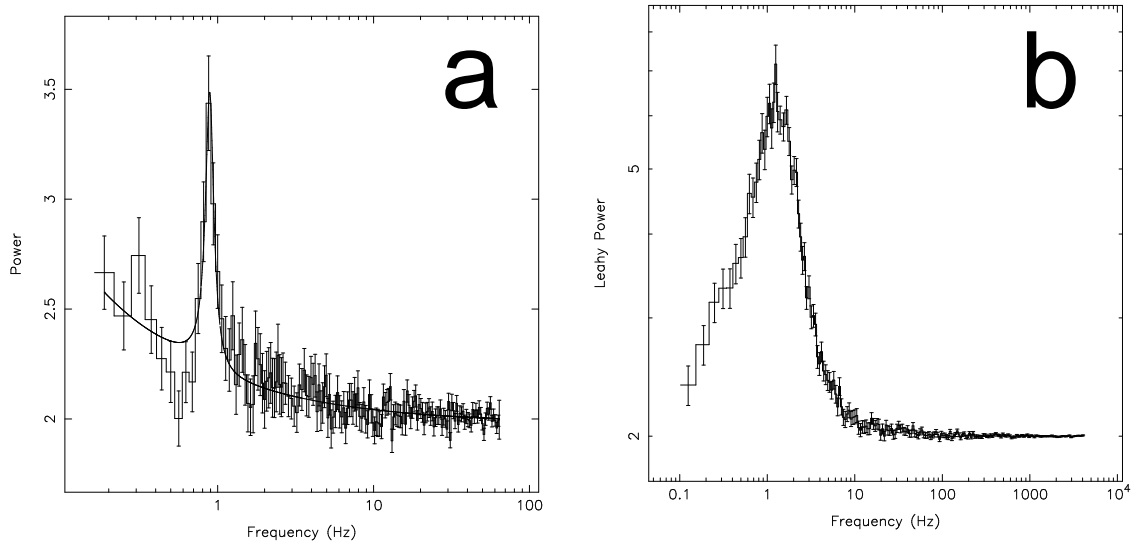}
\end{center}
\caption{The 1 Hz QPOs detected in the neutron star X-ray dipper 4U
1323--62 ({\it a}; Jonker et al. 1999) and in the millisecond X-ray
pulsar SAX J1808.4--3658 ({\it b}; van der Klis et al. 2000).
\label{fig:1HzQPO} }
\end{figure}

Low-frequency QPOs were also found in the neutron star X-ray dippers
but with frequencies near 1 Hz (Jonker et al. 1999, 2000a; Homan et
al. 1999; Fig.~\ref{fig:1HzQPO}{\it a}).  These QPOs are most likely
related to the high-inclination of the X-ray dippers and could be due
to matter at a certain disk radius coming periodically into the line
of sight (Jonker et al. 1999). The properties of these 1 Hz QPOs
suggest that they form a separate class of QPOs and that they are not
related to the other low-frequency QPOs observed in the Z and atoll
sources (Jonker et al. 1999, 2000a; Homan et al. 1999).

Recently, in several (but not all) RXTE observations taken of the
millisecond X-ray pulsar during a long-period of low-level activity in
early 2000, very strong flaring was observed which showed up as a very
strong (up to 100\% rms) 1 Hz QPO in the power spectrum (van der Klis
et al. 2000; Fig.~\ref{fig:1HzQPO}{\it b}). However, the differences
between this QPO and the 1 Hz QPOs in the X-ray dippers (i.e., the
coherence, the strength [100\% rms versus $\sim$10\% rms], and the
likely low-inclination of SAX J1808.4--3658) suggest that these QPOs
are caused by different physical mechanisms despite the similar
frequencies. Similar strong flaring behavior has not been observed in
the other atoll sources, but it might be present in these systems if
they reach similar low mass accretion rates as were observed for SAX
J1808.4--3658. However, it is also possible that this flaring is
related to the fact that SAX J1808.4--3658 is a millisecond X-ray
pulsar and it might then not be observable in the other non-pulsating
atoll sources.

\section*{A NEW VIEW ON NEUTRON STAR LOW-MASS X-RAY BINARIES}

The new obtained results with the RXTE satellite demonstrate that the
timing behavior of the neutron star LMXBs is far more complex than
previously thought. However, some general trends are already becoming
visible. First of all, with RXTE all types of QPOs previously thought
to be only present in the Z sources have now also been observed in the
atoll sources. This has profound implications for the models for these
QPOs and for our understanding of the differences and the similarities
between Z and atoll sources.  The Z and the atoll sources are likely
to be much more similar than previously suspected and the individual
sources only differ in detail from each other (the shape in the
color-color diagram, the exact detailed shape of the power spectra and
its evolution).  The general behavior of both the Z sources and the
atoll sources as suggested by these recent results can be summarized
as follows:

At low \mdot~the power spectra are dominated by strong band-limited
noise which follows approximately a power law at high frequencies but
breaks at a certain frequency below which the power spectra are
roughly flat. At frequencies above the break either a QPO or a broad
bump is present. At frequencies around approximately 100 Hz a broad
noise component is present (note that a detailed comparison of the Z
sources HFN and the atoll source 100 Hz noise has not yet been
performed which makes a conclusive statement about their connection
impossible). KHz QPOs might already be detectable in some sources, but
not in all. As \mdot~increases the frequency of the break and of the
QPO (the bumps evolve into true QPOs) increases and the strength of
both components decreases. The 100 Hz component does not change
significantly (Wijnands \& van der Klis 1998b; van Straaten et
al. 2000; di Salvo et al. 2000), although it might become more
coherent (van Straaten et al. 2000). If not yet already present, kHz
QPOs appear and their frequencies increase with \mdot.  The
band-limited noise, the low-frequency QPOs, and the kHz QPOs gradually
disappear as \mdot~increases further and the power spectra become
increasingly dominated by a very low frequency noise component
following a power law. QPOs near 5--7 Hz appear when the sources are
at very high \mdot, but these high \mdot~levels might not be reached
in all systems. If \mdot~increases even further either these QPOs
disappear or first increase in frequency to about 20 Hz before
disappearing.

Possible exceptions to this general picture are the X-ray dippers and
the millisecond X-ray pulsar. In the X-ray dippers, the general
Z/atoll source behavior is likely to be distorted by the fact that we
see these systems almost edge-on (e.g., Jonker et al. 1999). Different
timing phenomena can occur at such high inclination (i.e., the 1 Hz
QPOs), which are unrelated to the timing phenomena seen in the other
low-inclination NS systems. The violent flaring in SAX J1808.4--3658
might be unique to this source and related to the fact that this
source is the only known source of its kind, or perhaps similar
phenomena might be detected in the other atoll sources when they are
accreting at similarly low levels.

A fundamental question is why the atoll sources and the Z sources
resemble each other in the way they do, despite their significant
different luminosities. It is still likely that the mass accretion
rate is significantly higher in the Z sources compared to the atoll
sources, but it is unclear why the timing properties should be so
similar.  These similarities make it also increasingly unclear whether
a difference in magnetic field strength is really present between the
Z and the atoll sources.  The source which might give more insight
into the Z and atoll source connection is the enigmatic source Cir
X-1. With EXOSAT it was observed during episodes of low-level X-ray
activity and during that time it resembled the atoll sources
(Oosterbroek et al. 1995). However, from the start of the RXTE
mission, this source has continuously exhibited much higher flux
levels during which it resembles the Z sources, both in the track
traced out in the CD and in the timing phenomena (Shirey et
al. 1999). It seems that this source can alternate between atoll-like
states and Z-like states. Interestingly, the X-ray activity of Cir X-1
is steadily decreasing in the last few months which might indicate
that the source is moving back to a much lower X-ray activity.

\section*{BLACK-HOLE CANDIDATES}

\begin{figure}[t]
\begin{center}
\includegraphics[width=180mm]{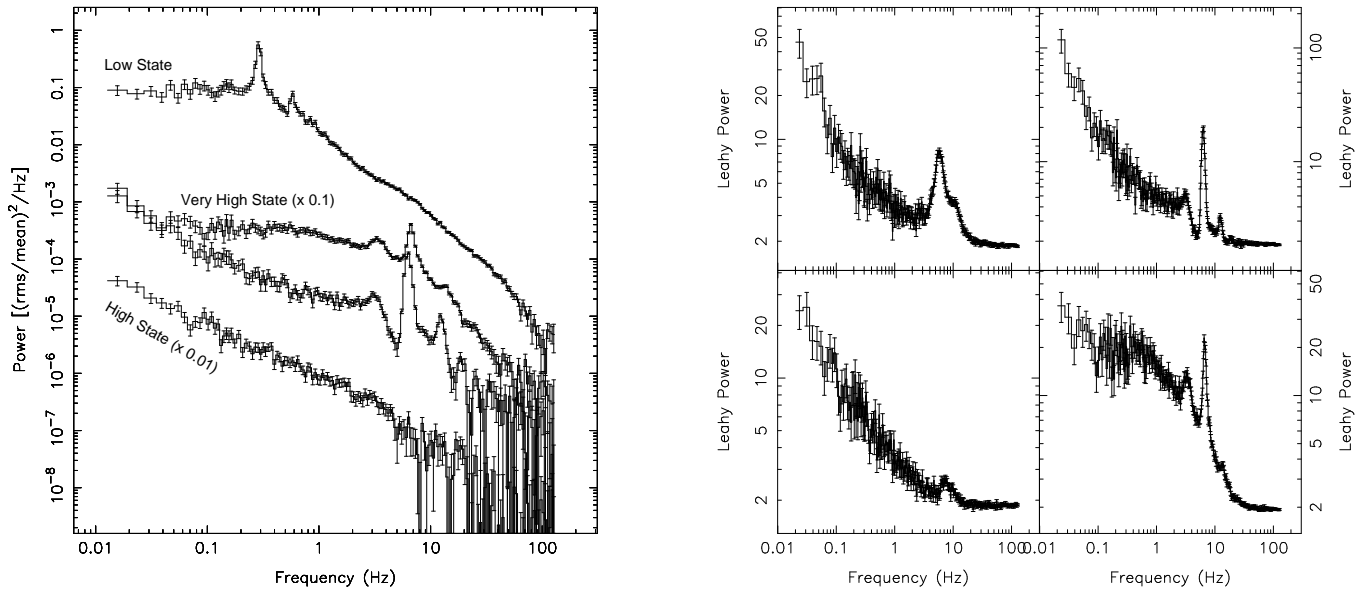}
\end{center}
\caption{Examples of the power spectra observed in the different
canonical BHC states ({\it left}) and the different types of VHS QPOs
observed in the recent BHC X-ray transient XTE J1550--564 ({\it
right}; Homan et al. 2000).  \label{fig:BHC}}
\end{figure}

Before RXTE, the 'standard' picture for black-hole candidates (BHCs)
was a simple, one-dimensional view: the changes in the X-ray spectra
and the rapid X-ray variability are caused by changes in \mdot~(Tanaka
\& Lewin 1995; van der Klis 1995). In the BHC low-state (LS), \mdot~is
low, the spectra are hard, and the power spectra are dominated by a
very strong (20\%--50\% rms amplitude) band-limited noise which
follows approximately a power law with index 1 at high frequencies but
below a certain frequency (\nubreak) the power spectrum becomes
roughly flat (Fig.~\ref{fig:BHC} {\it left}). Above \nubreak~a broad
bump or QPO is present, although also QPOs with similar frequencies
as the break frequency could be present (see Fig.\ref{fig:BHC} {\it
left}). In some sources a second break is visible in the power
spectrum above which the power law index increases to about
$\sim$2. In the high state (HS), \mdot~is higher, the spectra are much
softer, and in the power spectra only a weak (a few percent) power law
noise component is present. In the very high state (VHS), \mdot~is the
highest, the spectra are harder but not as hard as in the LS, and in
the power spectra noise is present similar to the weak HS noise or the
LS band-limited noise, although weaker (1\%--15\% rms).  Above
\nubreak, QPOs near 6 Hz are detected sometimes with a complex
harmonic structure.

With the many observations performed with RXTE on BHCs, the behavior
of BHCs turned out to be much more complex than previously
thought. First of all, RXTE has expanded the range of frequencies at
which the BHCs show variability up to 300 Hz (Remillard et
al. 1999a,b; Cui et al. 2000; Homan et al. 2000). The nature of these
BHC high-frequency QPOs is unclear, although recent evidence suggests
that they might be related to the lower-frequency peak of the twin kHz
QPOs in the neutron star systems (Psaltis et al. 1999). Also, with
RXTE in many BHCs the VHS QPOs have now been found (e.g., Remillard et
al. 1999b; Borozdin \& Trudolyubov 2000; Cui et al. 2000; Homan et
al. 2000; Revnivtsev et al. 2000), indicating that the VHS QPOs are a
common feature of BHCs. The phenomenology of these QPOs is very
complex and they are observed during different luminosity levels (at
levels significantly below the highest observed luminosities, i.e.,
not only during the VHS but at states intermediate between the LS and
the HS). An example of the complex structure of these QPOs is shown in
Fig.~\ref{fig:BHC} ({\it right}). In this figure different
low-frequency QPOs are shown which were observed during different
observations of the new BHC XTE J1550--564 (Homan et al. 2000).  The
interrelationships of the QPOs in this source are not fully
understood. The relationship between the QPOs observed in different
BHCs is even less understood.  However, from a detailed study of the
state behavior in XTE J1550--564 it is clear that the one-dimensional
picture described above for the BHC states (depending only on \mdot)
does not hold in this particular source and another extra parameter is
needed to explain its behavior (Homan et al. 2000). Similar behavior
might also be observable for the other BHCs, however, not much
information about this is available at the moment. A detailed
discussion of this and the QPOs observed in the BHCs is beyond the
scope of this manuscript.

\section*{THE \Large{\nubreakbold-\nuqpobold}~\normalsize RELATION}

\begin{figure}[t]
\begin{center}
\includegraphics[width=180mm]{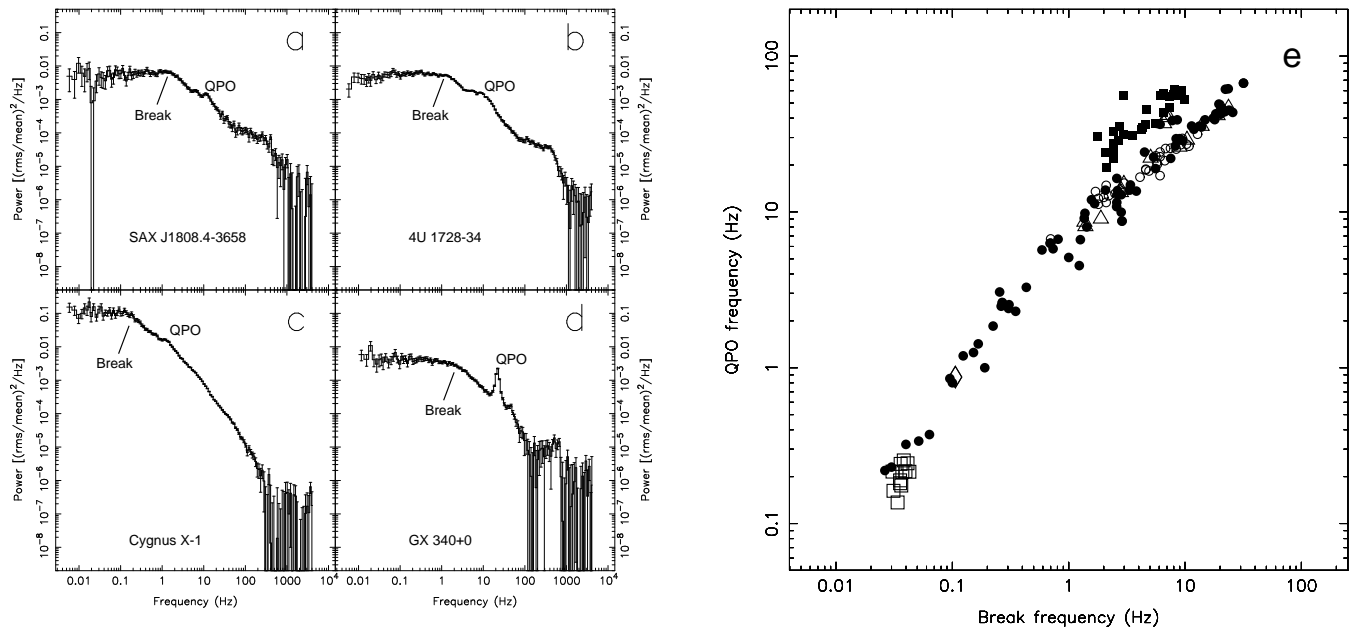}
\end{center}
\caption{Typical power spectra (see also Wijnands \& van der Klis
1999a) of the millisecond X-ray pulsar (SAX J1808.4--3658; {\it a}),
an atoll source (4U 1728--34; {\it b}), a BHC (Cyg X-1; {\it c}), and
a Z source (GX 340+0; {\it d}). The obtained correlation between the
frequency of the QPO versus the break frequency of the band-limited
noise is shown in {\it e}. The solid circles are the points for the
atoll sources (including the millisecond X-ray pulsar SAX
J1808.4--3658) and the BHCs used by Wijnands \& van der Klis (1999a);
the filled squares are their Z source data. The open symbols are new
data points obtained for several sources: GRO J0422+32 ({\it open
squares}; van der Hooft et al. 1999), SLX 1735--269 ({\it open
diamond}; Wijnands \& van der Klis 1999b), 4U 1728--34 ({\it open
triangles}; di Salvo et al. 2000), and 4U 0614+09 ({\it open circles};
van Straaten et al. 2000).
\label{fig:noise_correlation} }
\end{figure}

In a study to characterize the aperiodic variability of the
millisecond X-ray pulsar, Wijnands \& van der Klis (1998b) showed that
the power spectrum of this source is indistinguishable (besides the
pulsations) from that of the atoll sources when they are in the IS.
In Figures~\ref{fig:noise_correlation}{\it a}
and~\ref{fig:noise_correlation}{\it b} the power spectra are shown for
SAX J1808.4--3658 and a typical atoll source 4U 1728--34, respectively
(Wijnands \& van der Klis 1999a). For both sources the band-limited
noise and the bump above the break frequency are present, but so is an
extra noise component above 100 Hz. Wijnands \& van der Klis (1998b)
showed that this noise component is a common feature in the power
spectra of atoll sources at low mass accretion rates. The resemblance
between SAX J1808.4--3658 and 4U 1728--34 demonstrates that the
millisecond X-ray pulsar can be classified as an atoll source at low
\mdot~(i.e., it was in the IS; see Wijnands \& van der Klis 1998b). It
is unknown why for SAX J1808.4--3658 millisecond X-ray pulsations have
been detected and not for any other atoll source.

The striking similarities between SAX J1808.4--3658 and the atoll
sources, and the long-known similarities between the atoll sources in
the island state and BHC in the low state (see van der Klis 1995 and
references therein) initiated a detailed study to investigate these
similarities (Wijnands \& van der Klis 1999a). A typical power
spectrum of a BHC (i.e., Cygnus X-1) at low mass accretion rate is
shown in Figure~\ref{fig:noise_correlation}{\it c}. From this figure
it is clear that again the band-limited noise as well as the bump
above the break frequency are present. However, above the bump a
second break is present in the power spectrum above which the power
spectrum follows a power law with index of $\sim$2.  Wijnands \& van
der Klis (1999a) found a very good correlation between the frequency
of the QPO (or bump, but hereafter referred to as QPO; \nuqpo) and the
break-frequency (\nubreak; see Figure~\ref{fig:noise_correlation}{\it
e}). The most intriguing point is that the BHCs and the atoll sources
(including the millisecond X-ray pulsar) follow the same
correlation. This strongly suggest that in all those different source
types, the band-limited noise and the QPO are produced by the same
physical mechanism, which therefore cannot depend on the presence or
absence of either a small magnetosphere (which definitely is present
in SAX J1808.4--3658), a solid surface, or an event horizon. These
components then most likely originate somewhere in the accretion disk,
at a distance of at least several tens of kilometers from the central
compact object, outside a possible magnetosphere. However, the large
amplitudes of the band-limited noise (up to 50\% rms) exclude the
possibility that the emission carrying these fluctuations originates
this far out in the accretion disk, because most of the gravitational
energy of the accretion disk is released closer to the compact
object. A modulation of the accretion rate due to instabilities in the
flow in the region of the disk outside several 10 km from the compact
object is the most likely mechanism for generating the band-limited
noise and the QPO.

The question arises as to which fundamental properties of the source
determine the exact place of the source in \nubreak-\nuqpo~plot.  The
effect of the magnetic field of the compact object is probably small
in the region of the disk where the frequencies are determined. The
other physical parameters which can affect the accretion disk are the
mass and the spin of the central object, and \mdot. For a given source
the mass and spin do not change and thus probably only
\mdot~determines \nubreak~and \nuqpo. Usually the relation between the
frequencies and \mdot~is a positive one-to-one correlation (see, e.g.,
van der Klis 1995), however, it has been shown that this is not the
case for the millisecond X-ray pulsar (Wijnands \& van der Klis 1998b)
and the atoll source SLX 1735--269 (Wijnands \& van der Klis
1999b). The reason for this is as yet unknown, but it might be related
to the very low \mdot~observed in both sources. It however shows that
at least at these very low mass accretion rates another parameter
besides \mdot~is involved in determining the exact value of the
frequencies.  At higher accretion rates, when the frequencies are
positively correlated with \mdot, the difference between sources with
similar mass accretion rates would lie in the mass and/or the spin
rate of the compact object. This could explain why on average the BHCs
have smaller frequencies than the neutron star systems. However,
considerable overlap between these source types occurs, indicating
that the mass accretion rate differences dominate the frequencies.

\subsection*{Where do the Z sources fit?} 

When the brightest neutron star LMXBs (the Z sources; Hasinger \& van
der Klis 1989) are at low \mdot~(i.e., at the left end of the HB) the
power spectra are very similar to the one shown in
Figure~\ref{fig:noise_correlation}{\it d}. The LFN is present with the
HBO above the break. The shape of these power spectra resemble the
ones of the atoll sources and the BHCs at their lowest \mdot, although
the QPO is much more coherent for the Z sources. Also, for the Z
sources \nuqpo~and \nubreak~are correlated with each other, but they
do not follow the same correlation as the atoll sources and the BHCs
(see Figure~\ref{fig:noise_correlation}{\it e}; Wijnands \& van der
Klis 1999a).  Several arguments can explain why the Z sources follow a
different correlation. The most simple one is that the noise and/or
the QPOs in the Z sources are totally unrelated to any of the timing
phenomena in the atoll sources and the BHCs. However, the correlation
found between the frequencies of the kHz QPOs and the low-frequency
QPOs in both the Z sources and the atoll sources indicate that the
low-frequency QPOs in the atoll sources are likely to be related to
the HBOs in the Z sources (e.g., Ford \& van der Klis 1998; Psaltis et
al. 1999). So disregarding any common physical mechanism behind the
noise properties in the atoll and the Z sources seems premature. An
alternative explanation could be that the LFN in the Z sources is not
caused by the same physical mechanism as the HFN in the atoll sources
(see, e.g., Wijnands \& van der Klis 1999a). In the Z source Sco X-1
an extra noise component is present between \nubreak~ and
\nuqpo. Wijnands \& van der Klis (1999a) tentatively suggested that
this extra noise component could be related to the HFN in the atoll
sources. When plotting \nubreak~of this noise component versus \nuqpo,
Sco X-1 follows approximately the same relation as the atoll sources
and the BHCs (Wijnands \& van der Klis 1999a). However, Jonker et
al. (2000b) reported a similar noise component in the Z source GX
340+0. The frequency of this noise component was approximately half
the frequency of the HBO and could be its sub-harmonic. Identifying
this sub-harmonic with the bump or QPO in the atoll sources, GX 340+0
follows the same correlation as the atoll sources and the BHCs (Jonker
et al. 2000b). However, in this case GX 340+0 then does not follow the
correlations found by Psaltis et al. (1999) between the low-frequency
QPOs and the kHz QPOs.  At the moment it is unclear how the Z sources
fit in.

\subsection*{Cautionary statement}

The similarities between the neutron star systems and the BHCs at low
\mdot~(note that even some of the VHS QPOs of the BHCs fit into the
correlation which might point towards a physical connection between
these QPOs and neutron star 15--70 Hz QPOs; Wijnands \& van der Klis
1999a) are striking and the correlation between \nuqpo~and \nubreak~is
very suggestive of one underlying mechanism for the noise components
in both types of systems. But as can been seen in
Figure~\ref{fig:noise_correlation}{\it e}, intrinsic scatter is
present. The principal reason for the scatter could be the complex
structure of the QPO. Considerable substructure, usually below the
main QPO, is often present (see, e.g,
Figure~\ref{fig:noise_correlation}{\it a} and {\it b}). Moreover, in
several sources two QPOs were detected, harmonically related to each
other, or simultaneously QPOs and bumps were present (Wijnands \& van
der Klis 1999a). It is possible that in the other sources also extra
harmonics or bumps are present, which are incorporated into the single
Lorentzian used to fit the QPO. This would result in a measured QPO
frequency that is slightly shifted from the correct value, producing
the scatter.  When in the power spectra multiple QPOs, bumps, or even
multiple breaks are present, it is very difficult to identify which
QPO or which break to use for the correlation. Correlations as the
\nubreak-\nuqpo~correlation and the one found by Psaltis et al. (1999)
might be severely biased towards the simplest systems because the more
complex sources are not used in the correlations, resulting in an over
simplification of the timing behavior of X-ray binaries.

\subsection*{Can we distinguish between neutron star systems and BHCs?}

The above similarities between neutron star systems and BHCs make it
hard to distinguish these systems from each other and when an X-ray
binary does not exhibit X-ray pulsations or type I X-ray bursts, it is
usually hard to determine the nature of the compact object solely from
its rapid X-ray variability. However, a possible difference between
neutron stars and black-holes might be present in the rapid X-ray
variability above 100 Hz. First of all, so far no BHC has shown two
simultaneous kHz QPOs, although single QPOs up to approximately 300 Hz
have been detected in BHCs (e.g., Remillard et al. 1999a, b; Cui et
al. 2000; Homan et al. 2000).  The presence of two simultaneous kHz
QPOs strongly suggest the presence of a neutron star in the
system. But at low mass accretion rates, neither the neutron star kHz
QPOs (e.g., M\'endez et al. 1997) nor the black-hole high-frequency
QPOs (Remillard et al. 1999a,b; Homan et al. 2000) have been
detected. A possible distinction between the NS systems and BHCs at
these low accretion rates might still be possible on the basis of the
broad-band noise properties above approximately 100 Hz. As discussed
above, the neutron star systems display an extra noise component in
this frequency range, but in the BHCs this noise component is absent
(although van Straaten et al. [2000] tentatively suggested on the
basis of the sometimes high coherence of this noise component that it
might be related to the $>$100 Hz QPOs seen in BHCs).  Instead of this
100 Hz noise component the power spectra of the BHCs show a second
break above which the power spectra follow a steep (index $\sim2$)
power law. This difference in the high-frequency power spectra of BHCs
and neutron star systems was recently emphasized by Sunyaev and
Revnivtsev (2000) and it might turn out to be a useful tool to
determine the nature of the compact object in X-ray transients which
do not exhibit type I X-ray bursts nor pulsations.

\section*{CONCLUSIONS}

With the launch of RXTE a new area has begun in the study of the
X-ray variability of X-ray binaries. Although most of the attention
has been focused on the QPOs observed above 100 Hz,  a lot of new
and exciting phenomena have been discovered at lower frequencies. These
low-frequency phenomena seriously challenge the pre-RXTE view of X-ray
binaries. It will be a challenge to construct a new consistent view of
these systems which can both explain the striking similarities but
also the significant differences among the different source types.

\section*{ACKNOWLEDGEMENTS}

I would like to thank Luc\'\i a M.  Franco, Mariano M\'endez, Peter
Jonker, and Jeroen Homan for carefully reading this manuscript and
providing useful comments that helped to improved it. I would also
like to thank Jeroen Homan for help in preparing Figure~\ref{fig:BHC},
Tiziana di Salvo for sending me here results on 4U 1728--34 before
publications, and Frank van der Hooft for providing the data points
for GRO J0422+32.  This work was supported by NASA through Chandra
Postdoctoral Fellowship grant no. PF9-10010 awarded by CXC, operated
by SAO for NASA under contract NAS8-39073.  This research has made use
of data obtained through the HEASARC Online Service, provided by the
NASA/GSFC.

\end{document}